\documentclass[twocolumn,showpacs,preprintnumbers,amsmath,amssymb]{revtex4}
\usepackage{graphicx}
\usepackage{color}
\usepackage{subfigure}

\begin{document}

\title{Diffusion of two molecular species in a crowded environment: theory and experiments}
\author{D. Fanelli$^{1}$, A. J.~McKane$^{2}$, G. Pompili$^{1}$, B. Tiribilli$^{3}$, M. Vassalli$^{4}$, T. Biancalani$^{2}$}
\affiliation{$^{1}$Dipartimento di Energetica ``S. Stecco'' and INFN, 
University of Florence, Via S. Marta 3, 50139 Florence, Italy \\
$^{2}$Theoretical Physics Division, School of Physics and Astronomy, 
University of Manchester, Manchester M13 9PL, U.K. \\
$^{3}$CNR - ISC Sezione di Firenze, Via Madonna del Piano 10, 50019 Sesto 
Fiorentino, Florence, Italy \\
$^{4}$Institute of Biophysics CNR, 16149 Genoa, Italy  
}

\date{\today}

\begin{abstract}
Diffusion of a two component fluid is studied in the framework of differential
equations, but where these equations are systematically derived from a 
well-defined microscopic model. The model has a finite carrying capacity 
imposed upon it at the mesoscopic level and this is shown to lead to non-linear
cross diffusion terms that modify the conventional Fickean picture. After 
reviewing the derivation of the model, the experiments carried out to test the 
model are described. It is found that it can adequately explain the dynamics 
of two dense ink drops simultaneously evolving in a container filled with 
water. The experiment shows that molecular crowding results in the formation 
of a dynamical barrier that prevents the mixing of the drops. This phenomenon 
is successfully captured by the model. This suggests that the proposed model 
can be justifiably viewed as a generalization of standard diffusion to a 
multispecies setting, where crowding and steric interferences are taken into 
account.
\end{abstract}

\pacs{05.60.Cd, 05.40.-a, 87.15.Vv}

\maketitle

\section{Introduction}
\label{I:intro}
Diffusion is a key process in nature which describes the spread of particles
from regions of high density to regions of low density. As is usually the 
case, the first quantitative approach to the study of mass transport via 
diffusion was phenomenological, by Fick in the middle of the 19$^{\rm th}$ 
century. His first law relates the diffusive flux $\mathbf{J}({\mathbf{x}},t)$ 
to the concentration $\phi(\mathbf{x},t)$ at position $\mathbf{x}$ and 
time $t$, and mathematically amounts to the relation 
\begin{equation}
\label{1legge}
\mathbf{J}= -D \boldsymbol{\nabla} \phi,   
\end{equation}
where $D$ is a constant. Equation (\ref{1legge}), together with mass 
conservation, yields a diffusion equation for the concentration $\phi$:
\begin{equation}
\label{diff_eq}
\frac{\partial \phi}{\partial t} = D \nabla^2 \phi,   
\end{equation}
which is also referred to as Fick's second law~\cite{diffBook}. 

A microscopic justification for the diffusion equation followed some time 
later: Einstein~\cite{Einstein} and Smoluchowski~\cite{Smolouchowski} 
independently proposed a physical interpretation of the phenomenon, although 
based on the experimental work of Brown~\cite{Brown} almost a century earlier. 
They showed how the diffusion constant, $D$, can be related to measurable 
physical quantities such as viscosity and temperature. In addition, the famous 
relation between the mean square displacement 
$\langle \langle x^2 \rangle \rangle$ and the time elapsed $t$, 
$\langle \langle x^2 \rangle \rangle = 2 D t$, can be derived. Here double 
angle brackets indicate cumulants. The unintuitive, but now universally 
accepted, conclusion is that the average root-mean-square displacement of a 
Brownian particle grows with the square root of time. 

Despite the fact that the above picture, and Fickean diffusion in particular,
is generally appropriate when describing the spontaneous spatial rearrangement 
of particles in suspension, deviations are expected to occur in various 
situations of interest~\cite{bouchaud, metzler, zaslawsky}, e.g.~if obstacles 
are present or when different, and so distinguishable, species are sharing the 
available space at high concentration. A vast literature reports on these 
anomalies (see~\cite{bouchaud} and references therein); they are particularly 
important in applications to molecular biology, where a large ensemble of 
microscopic entities (e.g. proteins, macromolecules) are densely packed in 
space. This crowding effect can for example force the molecules populating the 
cells to behave in radically different ways than in test-tube 
assays~\cite{weiss, banks, dix, marcos}. For this reason, the study of 
biochemical processes under realistically crowded conditions is absolutely 
central to the understanding of cellular environments. These operating 
conditions are in fact ubiquitous in cells, and crowding could be an essential 
ingredient for an efficient implementation of the metabolic machinery. 

When crowding occurs --- a macroscopic effect reflecting the microscopic 
competition for the available spatial resources --- the mean-square 
displacement of particles is seen to scale in a sub-diffusive fashion. This 
empirical fact has so far being described by resorting to different 
fitting strategies for the experimental data. The most popular of these is to 
hypothesize a power law scaling for the variances~\cite{benavraham}. Fractional 
diffusion equations have also been proposed to explain the emergence of 
the scaling invoked~\cite{zaslawsky}, but the approach frequently appears to 
be a mathematically sound expedient which reproduces the observed behavior, but 
without a firm physical basis. 

Interestingly, practically all of the life science applications focusing on the 
formation of spatial and temporal patterns, assume reaction-diffusion schemes 
inspired by the mathematical paradigms of single species diffusion.  As such, 
the underlying partial differential equations have only a simple Laplacian 
term, and no accommodation is made for nonlinear corrections and/or couplings 
that need to be introduced if obstacles, or more than one species, are present 
at high concentration. 

By contrast, in an earlier work~\cite{FanelliMckane} a modification of Fick's 
law was proposed which was based on a detailed microscopic theory, which 
effectively encapsulated the essential aspects of limited spatial resources. 
This first-principles procedure represents a systematic approach to the 
modeling of biological, biochemical and related systems; it begins from an 
individual-based description of the interaction between the microscopic 
elements which make up the system. The competition for the available resources
leads to a modified (deterministic) diffusive behavior. More specifically, 
cross-diffusive terms appear which link multiple diffusing components and 
which modify the standard Laplacian term, a relic of Fick's law. The finite 
carrying capacity of each microscopic spatial patch is the physically 
motivated assumption at the heart of this extended theory of diffusion. The
theory has been used in the study of reaction-diffusion systems which 
are relevant to applications in the life 
sciences~\cite{BiancalaniDipattiFanelli,fanelliciancidipatti}. Other models
have been proposed which, starting from microscopic rules, lead to modified
macroscopic diffusion equations~\cite{Landman}.

This paper discusses an experimental investigation of the generalized 
model of diffusion presented in~\cite{FanelliMckane}. To this end, we will 
study the diffusion of ink drops located in the same spatial reservoir. As we 
will show, the diffusion of an isolated drop is successfully described in 
terms of standard diffusion. The case of different types of ink drops 
simultaneously evolving is more difficult to describe: the diffusion comes 
to an halt when contiguous drops make contact. This regime, clearly 
non-diffusive, is explained by invoking the multispecies setting 
of~\cite{FanelliMckane}---an observation that we shall quantitatively 
substantiate by benchmarking measurements to simulations. 

The paper is organized as follows. The next section, Section II, is devoted 
to introducing the model and reviewing the main steps of the derivation given 
in~\cite{FanelliMckane}. In Section III we will discuss the experimental 
apparatus and, in Section IV focus on the evolution of a single ink drop 
inside a Petri dish containing water. The coevolution of two drops is
studied in Section V, where fits to the data allow the accuracy of the theory
to be assessed. Finally, in Section VI we will sum up and conclude. 

\section{The model}
\label{s:model}
In the following we shall review the derivation of the model in the simplified 
case where two species of molecules diffuse, sharing the same spatial 
reservoir. It is however worth emphasising that the analysis extends 
straightforwardly to the generalized setting where an arbitrary number of 
molecular species are present. Similarly, the analysis holds in an arbitrary 
number of dimensions, but below we will assume that we are working in the 
physically relevant case of two dimensions.

The generic microscopic system that we consider will then be constrained to
occupy a given area of two-dimensional space. We assume that this area is 
partitioned into a large number $\Omega$ of small patches or cells. Each of
these mesoscopic cells, labeled by $i$, is characterized by a finite carrying 
capacity: it can host up to $N$ particles, namely $n^{(1)}_i$ of type $A^{(1)}$,
$n^{(2)}_i$ of type $A^{(2)}$, and $v_i = N-n^{(1)}_i - n^{(2)}_i$ vacancies, 
hereafter denoted by $V$. The molecules are assumed to have no direct 
interaction. There is however an indirect interaction which arises from the
competition for the available spatial resources. The mobility of the molecules 
will be {\it de facto} impeded if the neighboring cells have no vacancies. 
More concretely, we shall assume that molecules move only to nearest-neighbor 
cells, and only if there is a vacancy that can be eventually filled. This 
mechanism translates into the following chemical equation
\begin{equation}
A^{(a)}_i + V_j \stackrel{\mu^{(a)}}{\longrightarrow} V_i + A^{(a)}_j , \ \ a=1,2,
\label{mig}
\end{equation}
where $i$ and $j$ label nearest-neighbor cells with $A^{(1)}_i, A^{(2)}_i$, 
and $V_i$ identifying the particles (including the vacancies) in cell $i$. The 
parameters $\mu^{(1)}$ and $\mu^{(2)}$ denote the associated reaction rates.  
The state of the system is then characterized by the number of $A^{(1)}$ and 
$A^{(2)}$ molecules in each cell; the number of vacancies obeying a trivial 
normalization condition. To make the notation compact, we introduce the vector
$\mathbf{n}=(\mathbf{n}_{1},\ldots,\mathbf{n}_{\Omega})$, where 
$\mathbf{n}_i=(n^{(1)}_i,n^{(2)}_i)$. The quantity
$T(\mathbf{n}'|\mathbf{n})$ represents the transition rate from state
$\mathbf{n}$, to another state $\mathbf{n}'$, compatible with the former.
The transition rates associated with the migration between nearest-neighbors, 
as controlled by the chemical reaction Eq.~(\ref{mig}), can be cast in the form
\begin{equation}
T(n^{(a)}_{i}-1,n^{(a)}_{j}+1|n^{(a)}_{i},n^{(a)}_{j}) =
\frac{\mu^{(a)}}{z\Omega}\frac{n^{(a)}_i}{N} \frac{v_j}{N}, \ a=1,2, 
\label{TRs}
\end{equation}
where $z$ is the number of nearest-neighbors that each cell has. In addition, 
and to shorten the formulae, in the transition rates we only show the 
dependence on those particles which are involved in the reaction. As discussed 
in~\cite{FanelliMckane}, it is the factor $v_j = N-n^{(1)}_{j}-n^{(2)}_{j}$,  
reflecting the finite carrying capacity of cell $j$, that will eventually lead 
to the macroscopic modification of Fick's law of diffusion.

We assume that the process is Markovian, so that the probability of the system
being in state $\mathbf{n}$ at time $t$, $P(\mathbf{n}, t)$, is given by the
master equation~\cite{vk}
\begin{equation}
\frac{dP(\mathbf{n},t)}{dt} = \sum_{\mathbf{n}'\neq\mathbf{n}} 
\left[ T(\mathbf{n}|\mathbf{n}')P(\mathbf{n}',t) 
- T(\mathbf{n}'|\mathbf{n})P(\mathbf{n},t) \right],
\label{master}
\end{equation}
where the allowed transitions are specified by Eq.~(\ref{TRs}). Starting from 
this microscopic, and hence inherently stochastic picture, one can recover 
the corresponding macroscopic, deterministic description. To do this we need to
determine the dynamical equations for the ensemble averages 
$\langle n^{(1)}_{i} \rangle$ and $\langle n^{(2)}_{i} \rangle$. This is achieved
by multiplying the master equation (\ref{master}) by $n^{(a)}_i$ and summing 
over all $\mathbf{n}$. After an algebraic manipulation which requires shifting 
some of the sums by $\pm 1$, one finds
\begin{eqnarray}
\frac{d\langle n^{(a)}_{i} \rangle}{dt} &=& \sum_{j \in i} 
\left[ \langle T(n^{(a)}_{i}+1,n^{(a)}_{j}-1|n^{(a)}_{i},n^{(a)}_{j}) \rangle 
\right. \nonumber \\
&-& \left. \langle T(n^{(a)}_{i}-1,n^{(a)}_{j}+1|n^{(a)}_{i},n^{(a)}_{j}) 
\rangle \right], 
\label{pre_macro}
\end{eqnarray}
where the notation $\sum_{j \in i}$ means that we are summing over all patches 
$j$ which are nearest-neighbors of patch $i$.

The averages in Eq.~(\ref{pre_macro}) are explicitly carried out by recalling 
the expression for the transition rates (\ref{TRs}) and then replacing the 
averages of products by the products of averages, a formal step which is 
legitimate in the continuum limit $N \to \infty$~\cite{vk}. Rescaling time 
by a factor $N\Omega$, gives~\cite{mck04}
\begin{eqnarray}
\frac{d\phi^{(1)}_i}{dt} &=& \mu^{(1)} \left[ \Delta \phi^{(1)}_{i} 
+ \phi^{(1)}_{i} \Delta \phi^{(2)}_{i} - \phi^{(2)}_{i} 
\Delta \phi^{(1)}_{i} \right],
\nonumber \\
\frac{d\phi^{(2)}_{i}}{dt} &=& \mu^{(2)} \left[ \Delta \phi^{(2)}_{i} 
+ \phi^{(2)}_{i} \Delta \phi^{(1)}_{i} - \phi^{(1)}_{i} 
\Delta \phi^{(2)}_{i} \right].
\nonumber \\
\label{macro}
\end{eqnarray}
In the above expression we have introduced the concentration 
$\phi^{(a)}_{i} = \lim_{N \to \infty} \frac{\langle n^{(a)}_{i} \rangle}{N}$. The
symbol $\Delta$ denotes the discrete Laplacian operator which reads
$\Delta f_{i} = (2/z) \sum_{j \in i} (f_{j} - f_{i})$. Finally, by taking the 
size of the cells to zero, and scaling the rates $\mu^{(a)}$ 
appropriately~\cite{mck04} to obtain the diffusion constants $D^{(a)}$, one 
finds the following partial differential equations for the continuum 
concentration $\phi^{(1)}(\textbf{x},t)$ and $\phi^{(2)}(\textbf{x},t)$:
\begin{eqnarray}
\frac{\partial \phi^{(1)}}{\partial t} &=& D^{(1)}\left[ \nabla^2 \phi^{(1)} 
+ \phi^{(1)} \nabla^2 \phi^{(2)} - \phi^{(2)} \nabla^2 \phi^{(1)} \right], 
\nonumber \\
\frac{\partial \phi^{(2)}}{\partial t} &=& D^{(2)} \left[ \nabla^2 \phi^{(2)}
+ \phi^{(2)} \nabla^2 \phi^{(1)} - \phi^{(1)} \nabla^2 \phi^{(2)} \right],
\label{pdes}
\end{eqnarray}
where $\nabla^2$ is the usual Laplacian.

The extra terms $\pm(\phi^{(1)}\nabla^2\phi^{(2)}-\phi^{(2)}\nabla^2 \phi^{(1)})$ 
in the diffusion equations (\ref{pdes}) arise directly from having included 
vacancies in the transition probabilities. These nonlinear modifications of the 
simple diffusion equation have therefore a specific microscopic origin, as 
opposed to various ad hoc suggestions that have sometimes been proposed. These
equations cannot be solved exactly; we have studied them numerically using both
the forward Euler method in time and a semi-implicit algorithm which exactly 
conserves the number of particles~\cite{Brugnano}, and checked that they agree. 
The explicit results given in this paper were obtained using the forward Euler 
method.  

Since the two types of molecules are separately conserved, $\phi^{(1)}$ and 
$\phi^{(2)}$ satisfy the conservation relations
$\partial \phi^{(1)}/\partial t + {\rm div}\mathbf{J}_{\phi^{(1)}} = 0$ and 
$\partial \phi^{(2)}/\partial t + {\rm div}\mathbf{J}_{\phi^{(2)}} = 0$. This
means that the fluxes $\mathbf{J}_{\phi^{(1)}}$ and $\mathbf{J}_{\phi^{(2)}}$
satisfy
\begin{eqnarray}
\mathbf{J}_{\phi^{(1)}} &=& - D^{(1)} \left( \boldsymbol{\nabla} \phi^{(1)} + 
\phi^{(1)} \boldsymbol{\nabla} \phi^{(2)} - 
\phi^{(2)} \boldsymbol{\nabla} \phi^{(1)} \right), 
\nonumber \\
\mathbf{J}_{\phi^{(2)}} &=& - D^{(2)} \left( \boldsymbol{\nabla} \phi^{(2)} + 
\phi^{(2)} \boldsymbol{\nabla} \phi^{(1)} - 
\phi^{(1)}\boldsymbol{\nabla} \phi^{(2)} \right).
\label{modified_Ficks}
\end{eqnarray}
The above relation generalizes Fick's first law (\ref{1legge}).

In~\cite{FanelliMckane} a quite complete analysis of Eqs.~(\ref{pdes}) is 
given in the simplified case when the two diffusion constants are equal. It
is useful to define $\Phi^{(a)} \equiv \int \phi^{(a)}\,d \mathbf{x}$, $a=1,2$, 
where $\Phi^{(a)}$ quantifies the amount of material of type $A^{(a)}$. The 
integrations are carried over the space where the co-evolving molecular species
are confined. In the simulations performed later, this will be a square domain 
of size $L$. It is therefore useful to introduce 
$\bar{\phi^{(a)}} = \Phi^{(a)}/L^2$. Here, $\bar{\phi^{(a)}}$  can be interpreted 
as a type of (non-dimensional) average number density. 

In this paper, our aim is to provide an experimental verification of the 
generalized diffusion equations as obtained above. To reach this goal, we have 
constructed a simple experimental apparatus that enables us to quantitatively 
track the evolution of ink drops in water, and specifically considered the 
case of interest where two drops are simultaneously present. The theory will 
be tested against direct measurements so as to assess its validity. The next
section is devoted to introducing the experimental set-up which we use.

\section{The experimental apparatus}
\label{s:expt}
The experiments that we will discuss in this paper are aimed at monitoring 
the progressive diffusion of ink drops when deposited in water. This has been
previously studied systemically only for a single drop~\cite{diff_ink}. We
will reproduce these results, but only as a prelude to the real case of 
interest: the interaction between two drops. In this Section we will briefly 
describe the experimental apparatus, as well as the procedure of data 
acquisition which we used.

\begin{figure}[tb]
\begin{center}
\includegraphics[width=7cm,height=8 cm ]{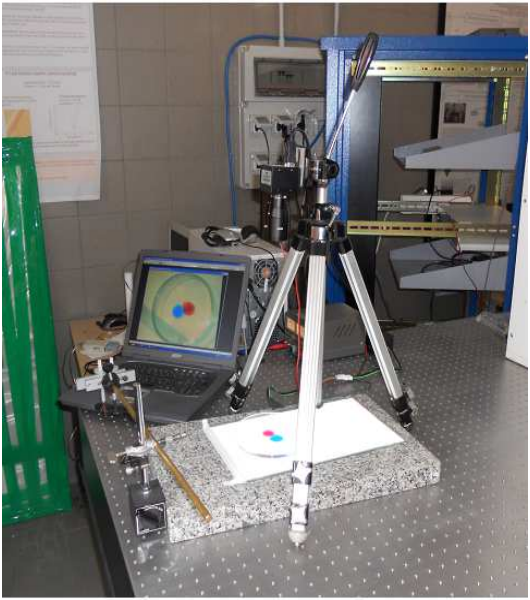} 
\end{center}
\caption{(Color online) The experimental set up. The camera is suspended 
above the Petri dish and  the system is illuminated from below by a light 
emitting capacitor. The simultaneous insertion of the two drops is carried 
out with a mechanical rod which holds two graduated injectors together. This 
device allows for the change of the distance between the drops at will, and 
in a completely reproducible fashion.}
\label{fig1}
\end{figure}

A Petri dish of diameter 14cm  was filled with a uniform 
layer (height 0.5cm) of water. The water was treated with a small amount of surfactant, to 
reduce the effects of surface tension. The container was placed on a 
horizontal surface, illuminated from below with a light emitting capacitor 
(LEC) produced by Ceelite~\cite{Ceelite} and distributed by Continua 
Light~\cite{CL}. This was the only source of light employed during the course 
of the experiment. Using an illumination in transmission strategy enabled us 
to substantially reduce the impact of spurious, undesirable shadows on the 
recorded images. The LEC panels were lightweight and extremely flexible; 
those in the experiment had dimensions of 21cm$\times$35cm. A color CCD camera 
(DFK 21AF04)~\cite{CCD} was placed 40cm above the container and controlled 
remotely via the computer, with an image of the sample being taken every 5 
seconds. A photograph of the experimental set-up is displayed in 
Fig.~\ref{fig1}. 

The solution was left to settle for several minutes so as to damp out any 
source of uncontrolled perturbations, such as local vortices or turbulent 
flows, that may develop when the water is put into the dish. Single, or 
multiple, drops of ink were then introduced with a graduated injector. The 
ink used in the experiments was of the Ecoline type, Talens 
brand~\cite{Talens}. Two specific colors were used: magenta and cyan. Upon 
injection, the ink formed a roundish drop which, under the action of gravity, 
fell to the bottom of the Petri dish. The system was therefore essentially two 
dimensional, the thickness of the ink layer being very small (of the order of 
tens of microns). The drop now started growing: the circular profile being 
preserved for isolated drops, and, in this case, the average drop radius 
progressively increased with time. By contrast, when two or more drops were 
present, mutual interference occurred that slowed down the diffusion, which 
was translated into distorted profiles. These latter profiles are analyzed in 
Sec.~V in the simplified case where just two drops were made to simultaneously 
evolve. In the next section we report on the experiments carried out on 
isolated drops.

\section{Diffusion of one isolated drop}
\label{s:one}
\begin{figure}[tb]
\begin{center}
\includegraphics[scale=0.5]{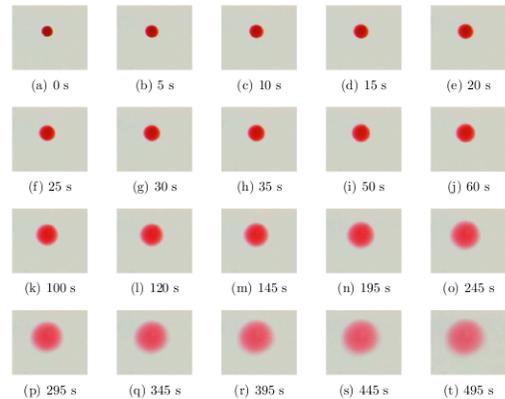} 
\end{center}
\caption{(Color online) Magenta drop: a gallery of images (a)-(t) is displayed 
at different times covering 495s with acquisition step of 5s. The images are 
treated so as to isolate the region of interest where the drop is localized. 
The experiments are carried out at a temperature of $17 \pm 1 ^{o}C$.    
}
\label{fig2}
\end{figure}

The simplest situation, and the one which we used to calibrate parameters 
associated with the ink, was when one isolated drop was allowed to form within 
the Petri dish~\cite{diff_ink}. The ink was injected at one specific location;
varying the amount of material that was introduced did not alter the fact that
it formed into a regular, almost two dimensional, circular domain. The region 
of interest in the recorded image was then cropped, and the operation repeated 
for all successive snapshots of the dynamics, resulting in the gallery of 
images displayed in Fig.~\ref{fig2}. A movie of the dynamics of the drop is 
attached as supplementary material.

The photographs that were obtained were then processed using the freely 
available ImageJ software~\cite{Image}. Depending on the color of the ink, a
specific filter was chosen, which maximized the corresponding adsorption and 
so enhanced the contrast of the drop. Thus, the green channel was employed 
when analyzing the magenta signal, while the red channel was used for the cyan 
drop. This choice simply follows from the subtractive theory of the CMYK color 
model~\cite{CMYK}.

Each image, converted into an intensity matrix, was normalized to the value 
of the maximum and then expressed in binary form. This was carried out by 
choosing a given (arbitrary --- see later discussion) threshold and setting 
pixels with an intensity lower than this threshold to 0, and those with an 
intensity above this threshold to 1. Each binary map was filtered to eliminate 
isolated pixels and small spurious features.  

\begin{figure}[tb]
\begin{center}
\includegraphics[width=9cm,height=7 cm ]{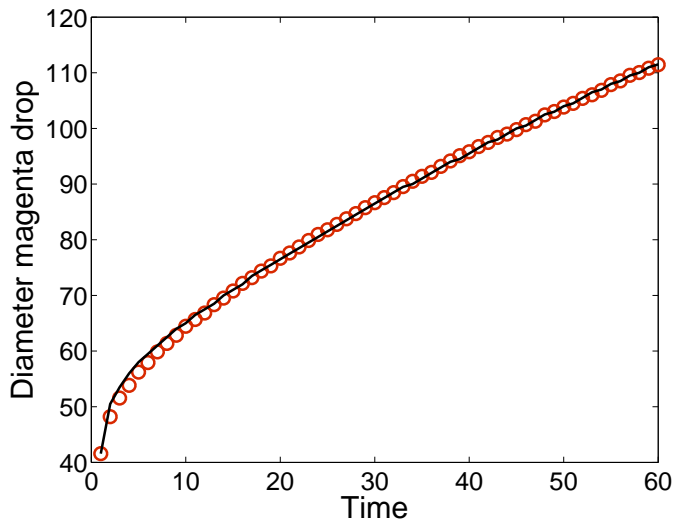} 
\end{center}
\caption{(Color online) The diameter $d_m$ of the magenta drop plotted versus 
time. The diameter is expressed in pixels ($1$ pixel $\simeq$ $0.33$mm). 
Snapshots are taken every 5s. Symbols refer to the experimental data, the 
diameter of the drops being calculated via an automatic procedure which 
transforms the two-dimensional intensity plot into a binary matrix. The solid 
line represents the solution of Eq.~(\ref{diff_eq}), where the diffusion 
coefficient $D_m$ and the relative threshold $\gamma_m$ are adjusted via a 
least square fitting procedure that maximizes the agreement between theory 
and experiments. In the simulations, at $t=0$, the drop is assumed to 
uniformly occupy a circular area. The radius of the circle (in pixels) is 
assigned so as to approximately match the size of the drop as seen in the 
first of the available experimental images. The simulations are performed 
over a square domain $L \times L$, with $L=150$ pixels. The best fit values 
are $D_m=0.7712$ pixels$^2$s$^{-1}$ and $\gamma_m=0.0675$.
}
\label{fig3}
\end{figure}

From this set of binary images, the diameter $d_m$ of the circular profile 
corresponding to the magenta drop was determined. For definiteness, we shall 
focus on the magenta drop to describe the procedure; the subscript $m$ denotes
magenta. The data are plotted as magenta colored circles in Fig.~\ref{fig3}.
This quantifies the growth of the drop, which was qualitatively displayed using 
the raw data in Fig.~\ref{fig2}. In Fig.~\ref{fig3}, the vertical axis is 
$d_m$, expressed in pixels. The conversion from pixels to physical units is 
straightforward, provided the resolution of the camera used is known (see 
caption of Fig.~\ref{fig3}).

We hypothesise that the observed dynamics is purely diffusive~\cite{diff_ink}. 
Motivated by this belief, we seek to establish a direct connection between 
the experimental data points, as displayed in Fig.~\ref{fig3}, and 
Eq.~(\ref{diff_eq}), which governs the dynamics of diffusive processes. It is 
worth emphasising that Eq.~(\ref{diff_eq}) governs the time evolution of the 
density of the diffusing species, while the experimental points recorded are 
calculated from intensity signals as registered by the camera. Comparison of
the profiles from the equation and from the data is therefore not 
straightforward, since the transformation from intensities to densities is 
highly non-linear. It is possible, however, to make a comparison at the level
of the (circular) domain obtained from data in the manner described above. The
idea is to numerically integrate Eq.~(\ref{diff_eq}) with the diffusion 
coefficient $D$ (here identified as $D_m$) being adjusted iteratively so as 
to maximize the agreement between the theoretical and experimental curves. This
is not the only free parameter: recall that the threshold imposed when 
performing the binary conversion was arbitrarily chosen. Once again, the
transformation between that threshold parameter and the one which delimits the
boundary of the drop in the numerical integrations of the equation will be 
highly non-linear. However, we do not need to characterize this mapping, and 
need only know that there is a one-to-one correspondence between them, using 
the threshold parameter in the numerical integrations as the other free 
parameter of the theory. We denote it by $\gamma_m$. Therefore the numerical
solution of Eq.~(\ref{diff_eq}) will contain two free parameters: $D_m$ and
$\gamma_m$.

We have referred to the solution of Eq.~(\ref{diff_eq}) as obtained from a
numerical integration. It is, of course, possible to solve Eq.~(\ref{diff_eq})
analytically, but we choose to find the solution numerically because we wish 
to use the same method in the case of two drops, and to test the methodology 
in the present case of a single drop. Since the corresponding Eqs.~(\ref{pdes})
are not exactly solvable, we are forced in the case of two drops to use 
numerical integration methods. 

The procedure is now that defined above: to iteratively adjust the two 
parameters $D_m$ and $\gamma_m$ so as to maximize the agreement between the 
theoretical and experimental curves. However, the experimental data gives the
radius $d_m$ of the circular domain, and we need to relate this to $\gamma_m$.
We choose to do this as follows: if the drop extends over a two-dimensional 
circular domain of radius $d_m$, then 
$\phi(x,y,t)/{\textrm{max}}_{x,y}(\phi(x,y,t)) < \gamma_m$ for each point 
$\mathbf{x}=(x,y)$ in the domain.

\begin{figure}[tb]
\begin{center}
\includegraphics[width=9cm,height=7 cm ]{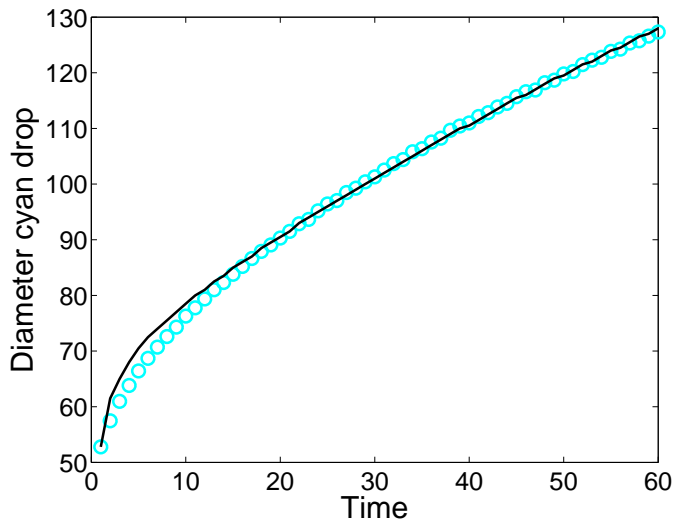} 
\end{center}
\caption{(Color online) The diameter $d_c$ of the cyan drop plotted against 
time. Symbols represent experimental data. The solid line represents the 
solution of Eq.~(\ref{diff_eq}), where the diffusion coefficient $D_c$ and 
the threshold parameter $\gamma_c$ are found via a fitting procedure. The 
initial condition in the simulations is set as described in the caption of 
Fig.~\ref{fig3}. The best fit values are $D_c=0.9459$ pixels$^2$s$^{-1}$  and 
$\gamma_c=0.0628$.
}
\label{fig4}
\end{figure}

The experimental data, represented by circles, are plotted as function of 
time in Fig.~\ref{fig3}. The solid line represents the solution of the 
diffusion equation (\ref{diff_eq}), with the diffusion constant $D_m$ and the 
threshold parameter $\gamma_m$ found using the iterative fitting scheme 
described above. The agreement is excellent, the evolution of the drop being, 
as expected, purely diffusive. Similar conclusions can be reached for the cyan 
drop. Experimental and theoretical computed diameters $d_c$ are compared in 
Fig.~\ref{fig4}. The diffusion constant $D_c$ and the threshold $\gamma_c$ for
cyan are found by analogy with that for magenta, that is, through the 
iterative fitting procedure described above.

We conclude that standard diffusion theory adequately explains the dynamics of 
isolated drops, as seen in the experiments that we carried out. A 
straightforward fitting strategy allowed us to calculate the associated 
diffusion constants for the two different colors of ink that we studied. These 
values will be assumed, and used as input, in the more complex scenario where 
two drops simultaneously evolve in the same Petri dish. We now go on to
discuss the corresponding analysis in this case.

\section{Diffusion of two drops: mutual interference and cross terms}
\label{s:two}
\begin{figure}[b]
\includegraphics[width=8cm,height=6 cm]{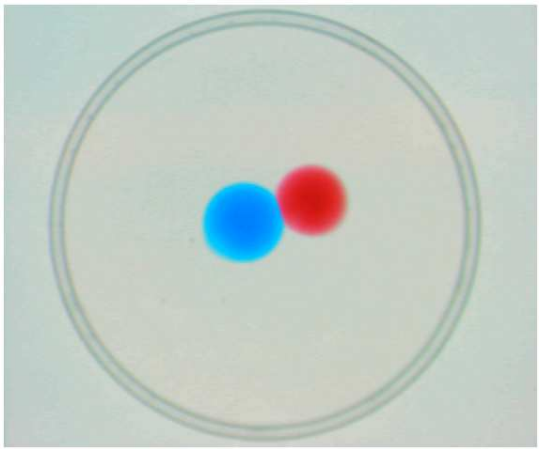} 
\caption{(Color online) The two drops in the Petri dish, some time after 
injection.   
}
\label{fig5}
\end{figure}

\begin{figure}[b]
\begin{center}
\includegraphics[scale=0.45]{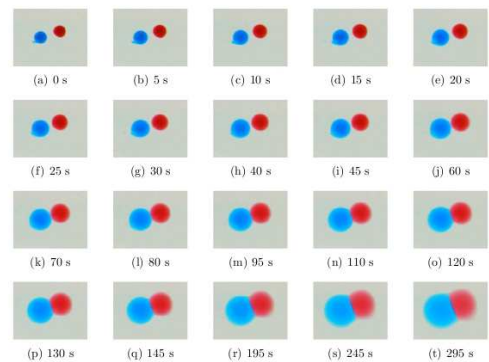} 
\end{center}
\caption{(Color online) A collection of images displaying the dynamics of 
the two drops at different times. The images cover a window of observation 
of 295s, with an acquisition step of 5s. A movie of the time evolution of 
the two drops is annexed as supplementary material.
}
\label{fig6}
\end{figure}

In this Section we discuss the diffusion of two ink drops, one magenta and the
other cyan, which were simultaneously inserted into the Petri dish described 
above. The drops were initially localized in different parts of the dish, and so
did not influence each other. Diffusion made them gradually expand, until they 
eventually came into contact, as shown in Fig.~\ref{fig5}. At this point, a 
sharp interface developed between the drops which acted as a barrier preventing 
the mixing of the two compounds. From then on, the constraints imposed by this
barrier distorted the previously circular profiles, as shown in 
Fig.~\ref{fig6}. We interpret this phenomenon as stemming from the 
competition of the molecular constituents for the finite available space. The 
saturation of microscopic spatial resources presumably results in locally 
congested configurations, that are responsible for the macroscopic features
seen in experiments.

\begin{figure}
\centering
\subfigure
{\resizebox{0.47\textwidth}{!}{ \includegraphics{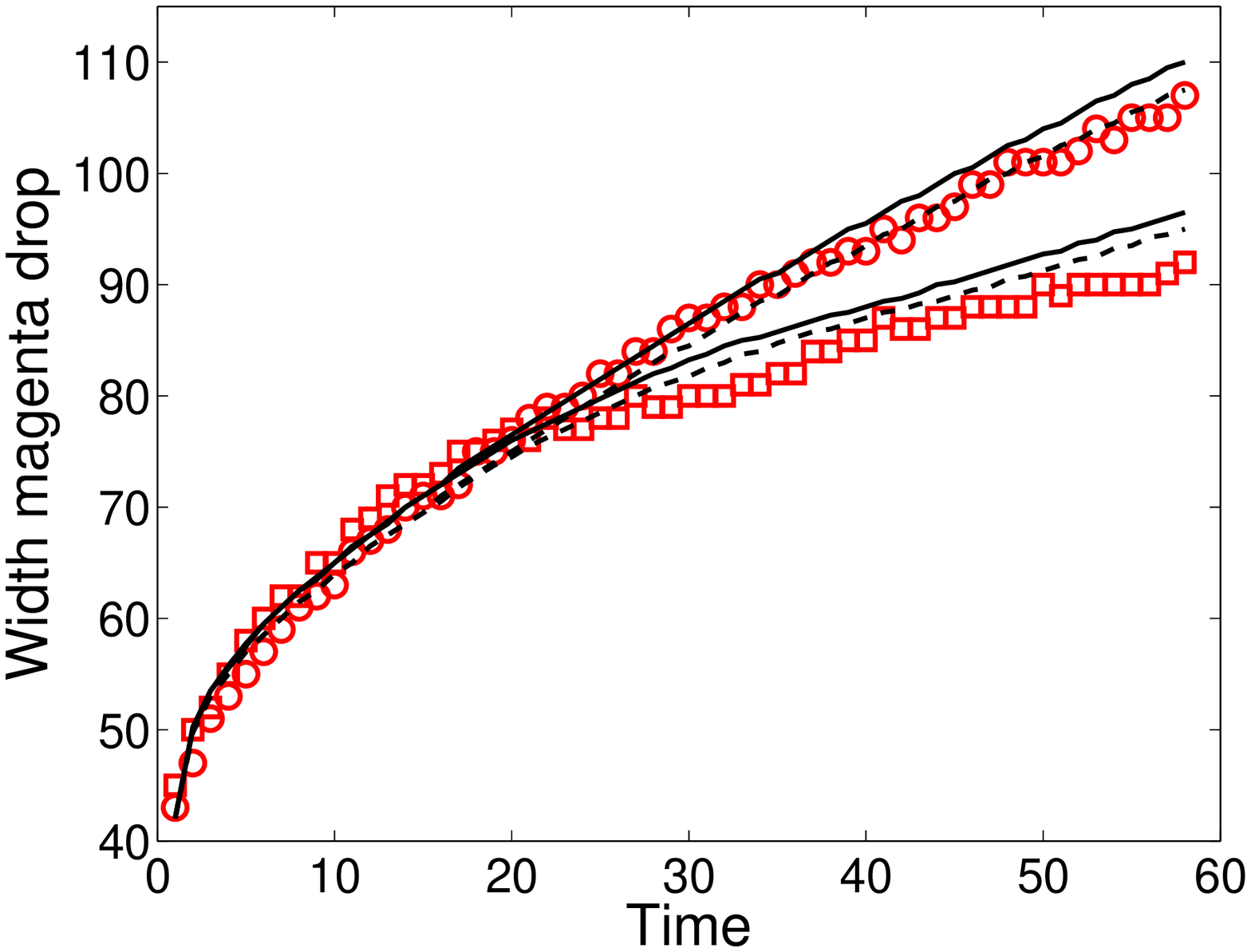}}}
\subfigure
{\resizebox{0.47\textwidth}{!}{ \includegraphics{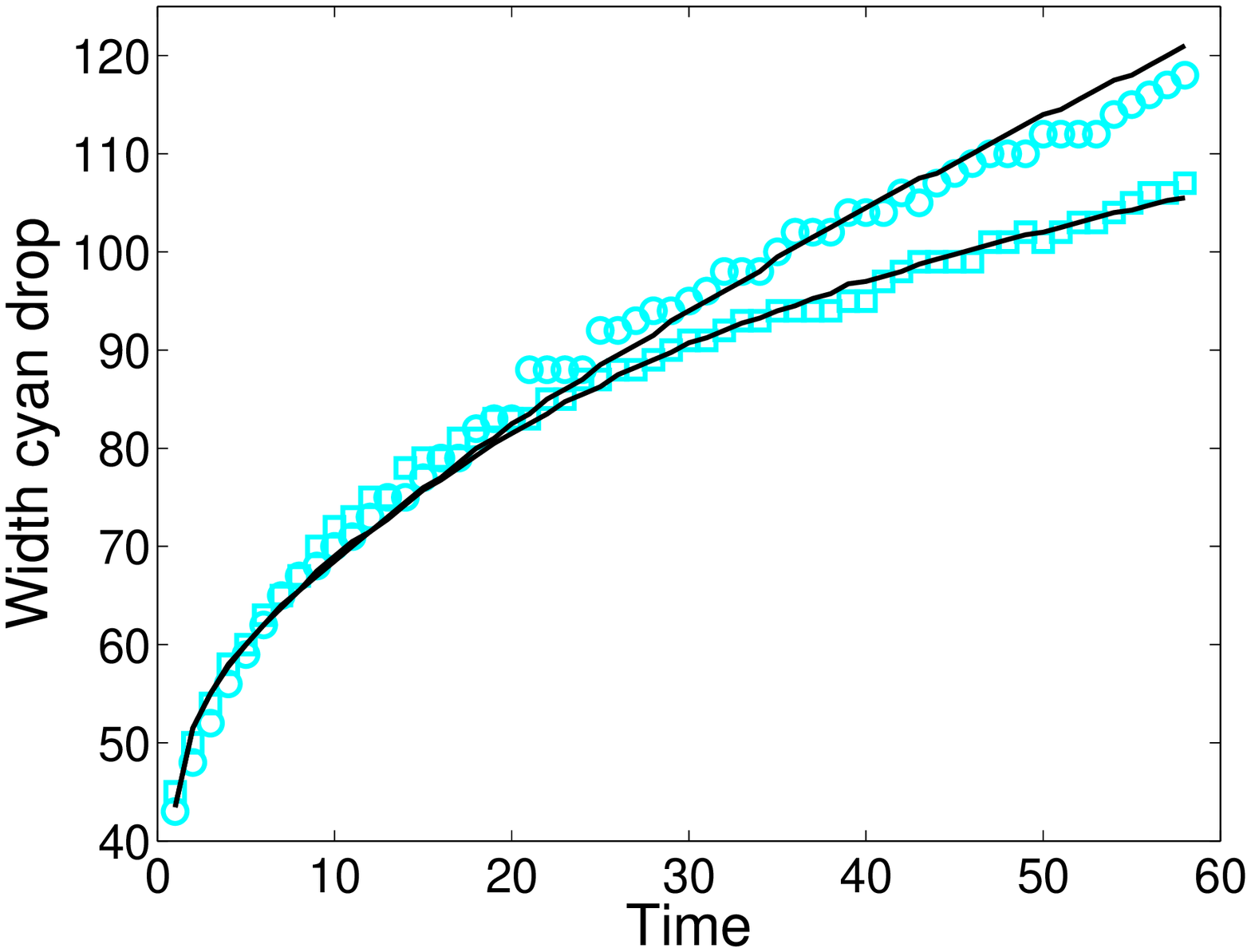}}}
\caption{(Color online) Upper panel: The diameters of the magenta drop are 
displayed versus time. Symbols refer to the experiments: open circles for 
the diameter of the drop in the $Y$ direction and squares for the diameters 
along $X$. The solid lines represent the numerical fit based on the 
generalized diffusion Eqs.~(\ref{pdes}). The dashed lines are the result of 
a generalized fitting strategy, described in the main text. Lower panel: As 
for the upper panel, but for the cyan drop. The simulations are performed 
over a square of linear size $L$, with $L=150$.
}
\label{fig7}
\end{figure}

To test this hypothesis we proceeded as follows. We defined an $X$-axis as the 
direction of the line connecting the centers of mass of the two drops at $t=0$.
A $Y$-axis was then defined as being orthogonal to the $X$-axis. The images 
recorded were then processed to measure the maximal extension of both of 
the drops, along the two independent directions $X$ and $Y$, at different 
times of acquisition of the signal. This therefore gave four quantities at 
each time $t$, two for the magenta drop and two for the cyan drop. The 
values obtained were found by filtering the recorded intensity map with 
an imposed threshold, following the strategy discussed above. This created
a binary representation of the two drops, the analysis returning four time 
series, two for each drop. The data are plotted in Fig.~\ref{fig7}: the upper 
panel displays the data corresponding to the magenta drop, the lower panel to
the cyan drop. In both panels, the circles represent the measurements carried 
out along the $Y$ direction, perpendicular to the line connecting the centers
of mass of the drops. The squares refer to the extension of the drops in the 
$X$ direction, along the line connecting the centers of mass.

The initial condition used in solving Eqs.~(\ref{pdes}) mimics that found in
the experiment. The drops, assumed to be approximately circular at $t=0$, are 
in fact assigned a specific size that is deduced from the experiments --- from
the first of the recorded images. Similarly, the relative separation between 
the drops is calculated in pixels and used as an input to the numerical 
simulations.

At relatively short times, before the magenta and cyan drops come into contact,
their diameters in the $X$ and $Y$ directions are approximately the same, their
shape being approximately circular. As already remarked, once the fronts 
collide ($t \simeq 100 s$, see Figs.~\ref{fig5} and \ref{fig6} panel $(m)$), the 
dynamics slows down at the interface of the two drops. The merging of the 
drops is consequently impeded along the direction of collision, a dynamical 
effect that contributes to reducing the growth of the drop width along the 
$X$ direction. By contrast, the diameter of both drops keeps on increasing 
along the $Y$ direction as if normal diffusion governed the dynamics. The 
symmetry of the system is therefore broken, as clearly shown in 
Fig.~\ref{fig7}.

To establish a bridge between the results of the experiments and the 
multispecies model of diffusion (\ref{pdes}), we used a fitting protocol 
based on the analysis developed in Sec.~\ref{s:one}. In particular, the 
coefficients $D^{(1)} \equiv D_c$ and $D^{(2)} \equiv D_m$ are obtained from
the single drop analysis, as are the respective thresholds $\gamma_c$ and 
$\gamma_m$. Two parameters are left undetermined, the rescaled normalization 
constants $\bar{\phi^{(1)}}$ and $\bar{\phi^{(2)}}$. These will serve as fitting
parameters, found by optimizing the agreement between theory and measurements. 
The difference between the two, that is minimized by the iterative fitting 
scheme, is the sum of the norm of relative deviations between the numerically 
computed and the experimentally measured diameters, along the two directions 
of observation, for both drops, at the same time. It should be emphasised that,
due to the nonlinearity introduced by the cross terms in Eq.~(\ref{pdes}), the 
normalization factors $\bar{\phi^{(1)}}$ and $\bar{\phi^{(2)}}$ do not trivially
cancel, as is the case if the linear diffusion equation (\ref{diff_eq}) is 
assumed to govern the dynamics. Larger values of $\bar{\phi^{(1)}}$ and 
$\bar{\phi^{(2)}}$, enhance the impact of cross terms and so magnify the 
deviations from the purely diffusive scenario. 

While $\bar{\phi^{(1)}}$ and $\bar{\phi^{(2)}}$ indirectly quantify the amount 
of material that was introduced into the Petri dish, they are also probably 
sensitive to the microscopic shape and steric properties of individual ink 
molecules, supposedly distinct for cyan and magenta samples. Moreover, 
$\bar{\phi^{(1)}}$ and $\bar{\phi^{(2)}}$ may encapsulate three-dimensional 
effects, which although assumed to be negligible in our treatment, are 
responsible for a slight deformation of the drops along the vertical, $z$ 
direction, an effect which, a priori, is different for the two drops. Based 
on these considerations, and despite the fact that care has been taken to keep 
the two drops as similar as possible at the time of injection into the dish, 
one should allow for different values of the unknown constants  
$\bar{\phi^{(1)}}$ and $\bar{\phi^{(2)}}$.

The result of the fitting procedure is reported in Fig.~\ref{fig7}. The 
fitting is first performed by only adjusting the average number densities  
$\bar{\phi}^{(1)}$ and $\bar{\phi}^{(2)}$; here cyan is the molecular species 
$1$ and magenta the molecular species $2$. The diffusivities, $D_c$ and $D_m$, 
as well as the thresholds, $\gamma_c$ and $\gamma_m$, are set to the values 
determined above when analyzing the single drop dynamics (see captions of 
Figs.~\ref{fig3} and \ref{fig4}). The time series for the widths of the 
two drops (a total of four independent curves) are hence simultaneously 
fitted by using only two control parameters. The best fit values are  
$\bar{\phi}^{(1)}= 0.3406$ and $\bar{\phi}^{(2)}=0.2126$.

The dashed lines in Fig.~\ref{fig7} are the result of a generalized fitting 
strategy, where the thresholds $\gamma_m$ and $\gamma_c$ are also allowed 
to vary. In this latter case, the fitted parameters result in 
$\bar{\phi}^{(1)}= 0.32815$, $\bar{\phi}^{(2)}=0.2233$, $\gamma_m=0.0727$ and 
$\gamma_c=0.0608$. In the case of the cyan drop, shown in the lower panel of
Fig.~\ref{fig7}, the dashed lines obtained from the generalized fitting 
strategy which uses four free parameters are not distinguishable from the 
solid lines, obtained when only two free parameters are used.

In summary, by adjusting two fitting parameters we are able to approximate 
four curves, representing the time evolution of the widths of the two drops 
along the reference directions $X$ and $Y$, with good accuracy. The slight 
deviation observed for the magenta drop is partially removed by allowing for 
the thresholds to change by a few percent, with respect to the values 
determined for the single drop case. The fact that $\gamma_m$ is modified by
the generalized fitting strategy has its origins in a cross-talk between color
channels. The cyan drop induces a small offset in the magenta channel, when 
the two drops are present simultaneously in the analyzed image. This effect is
responsible for the small change in $\gamma_m$, as compared to the value 
obtained for an isolated magenta drop.

The results obtained point to the validity of the model of multispecies 
diffusion described in Sec.~\ref{s:model}. The model is capable of mimicking 
the mutual interference between different chemical compounds confined to the 
same spatial reservoir, under crowded conditions, as seen in the experiments 
reported here.  

\section{Conclusion}

The problem of multispecies diffusion is of great importance in many areas of 
science. Particularly crucial is the field of biology, where distinct molecular
constituents, organized in families, populate (for instance) the cellular 
medium under crowded conditions. Among the models that have been proposed in 
the literature to analyze diffusion in a multi-component system, the one 
derived in~\cite{FanelliMckane} moves from a microscopic, hence inherently 
stochastic formulation of the problem, to macroscopic equations which 
incorporate the consequences of crowding. It does this by accounting for the 
effect of limited spatial resources, that seed an indirect competition between 
the different molecular species. The model of~\cite{FanelliMckane} therefore 
comprises of a set of (deterministic) differential equations which are fully 
justified from first principles, and which are also expected to apply in 
situations when large densities are present. In this paper we have tested this
model experimentally by investigating the diffusion of two ink drops which are
simultaneously evolving in a container filled with water. 

The experimental set up was calibrated with reference to the single drop case 
study, a simplified scenario that enabled us to measure the characteristic 
diffusion coefficients of the ink. When instead two drops are introduced into 
the container, a curious phenomenon takes place: after an initial evolution, 
which can be explained using the standard theory of diffusion, the fronts of 
the drops collide and a barrier is established at the interface which prevents 
mixing of the ink drops. This observation cannot be described by normal 
diffusion theory. The symmetry between the $X$ and $Y$ directions  (defined 
respectively to be the direction of the impact and that orthogonal to it) is 
in fact broken, an intriguing dynamical feature that we interpret as the 
macroscopic consequence of the competition for the microscopic spatial 
resources that takes place at the frontier between the two drops. To test the 
model proposed in~\cite{FanelliMckane} we have measured the widths of the 
drops, along the two independent $X$ and $Y$ directions and compared the 
experimental data points obtained to the numerical solution of the governing 
differential equations (\ref{pdes}). The numerical time series are compared
to the experiments by performing a constrained fitting, which employs a limited
number of free parameters. The fitting algorithm incorporates information from
an independent and preliminary analysis of the evolution of isolated ink drops 
in a similar environment. The agreement between theory and experiments is 
good, pointing to the validity of the mathematical model derived 
in~\cite{FanelliMckane}. We therefore conclude that this model accurately 
captures the dynamics of two, simultaneously diffusing, granular species, in 
the non-trivial situation where crowding occurs.

The results of the experiments reported in this paper, together with the 
derivation of the model and its numerical and analytical study, lead us to 
propose that it should be adopted in reaction-diffusion problems, where species
are known to be spatially dense, so abandoning the inappropriate Fickean 
approximation. This allows novel insights to be gained into several 
important mechanisms, for example Turing instabilities~\cite{turing1952} and 
the paradigmatic approach to pattern formation in biology~\cite{murray}. 
In~\cite{fanelliciancidipatti,kumar2011} it is shown that, due to non-Fickean 
cross-diffusion terms, the Turing instability also sets in when the activator 
diffuses faster then the inhibitor, at odds with the conventional, currently 
accepted, picture. We expect many other applications of this microscopically 
motivated approach to crowding will be explored in the future.

\acknowledgments
The work is supported by Ente Cassa di Risparmio di Firenze and the 
program PRIN2009. T.B. wishes to thank the EPSRC for partial support.

\end{document}